\newif\iflongbib
\longbibtrue
\documentclass[final,3p,times]{elsarticle}

\usepackage{amssymb}
\usepackage{amsmath}
\usepackage{amsthm}
\usepackage{mfirstuc}
\usepackage{xcolor}
\definecolor{linkc}{rgb}{0.1,0.1,.8}
\definecolor{darkgreen}{rgb}{0,0.5,0}
\definecolor{midblue}{rgb}{0,0,0.7}
\usepackage[bookmarks=false,colorlinks=true,citecolor=linkc,linkcolor=linkc,filecolor=linkc,urlcolor=linkc]{hyperref}
\makeatletter
\AtBeginDocument{%
  \pdfstringdefDisableCommands{%
    \def\cite#1{}%
    \def\citet#1{}%
    \def\citep#1{}%
    \def\@corref{}%
    \def\corref#1{}%
    \def\cortext#1#2{}%
  }%
}
\makeatother
\usepackage{url}
\usepackage{float}
\usepackage{subfigure}

\providecommand{\noopsort}[1]{}

\newcommand{\sysname}[1]{{\sc #1}}
\newcommand{\qsysname}[1]{{\sc \capitalisewords{\lowercase{#1}}}}

\journal{Journal of Symbolic Computation}

\begin{document}

\title{Symbolic Mathematical Computation 1965--1975:\\
The emergence of a discipline}


\author[WesternCS,Waterloo]{Robert M.~Corless\corref{cor1}}
\ead{rcorless@uwo.ca}

\author[Trinity]{Arthur Norman}
\ead{acn1@cam.ac.uk}

\author[Nebrija]{Tom\'as Recio}
\ead{trecio@nebrija.es}

\author[WesternHist]{William J.~Turkel}
\ead{william.j.turkel@gmail.com}

\author[Waterloo]{Stephen M.~Watt}
\ead{smwatt@uwaterloo.ca}

\affiliation[WesternCS]{organization={Department of Computer Science, Western University},
            city={London},
            country={Canada}}

\affiliation[Trinity]{organization={Trinity College},
            city={Cambridge},
            country={UK}}

\affiliation[Nebrija]{organization={Departamento de Matem\'aticas y F{\'\i}sica, Universidad Antonio de Nebrija},
            city={Madrid},
            country={Spain}}

\affiliation[WesternHist]{organization={Department of History, Western University},
            city={London},
            country={Canada}}

\affiliation[Waterloo]{organization={Cheriton School of Computer Science, University of Waterloo},
            city={Waterloo},
            country={Canada}}

\cortext[cor1]{Corresponding author}

\begin{abstract}
Today, symbolic mathematical computation is taken for granted as part of the scientific infrastructure, but it has not always been so.  
This paper provides a historical survey of the discipline’s formative decade, 1965--1975, viewed from a 50 year perspective.  
This span of years saw the evolution from a few specialized programs with naive algorithms
to integrated systems with substantial capabilities.
We highlight some of the important early figures in the field and the innovations upon which the current generation of systems and algorithms are built.
By revisiting a period unfamiliar to most current readers, this survey aims to shed light on once-pressing issues that are now largely resolved and to highlight how some of today’s challenges were recognized earlier than expected.
\end{abstract}


\maketitle
\section{Introduction}
The modern symbolic computation world is quite large in absolute terms, comprising on the order of a thousand active researchers and a million dedicated users, and even more people who use it occasionally\footnote{These numbers are not rigorous and may not be correct even within a factor of ten, because they depend on the exact definition of what it means to be an ``active researcher'' in the field, or a ``dedicated user.'' Still, by roughly counting the distinct attendees at various conferences, the number of authors of JSC articles, and the number of developers of various systems (not forgetting the large group of Symbolic Toolbox developers at The Mathworks), and acknowledging the 2018 public claim of ``100 million users of GeoGebra,'' which we asked the GeoGebra team about and they told us that as of 2025 the number of downloads was over 144 million,  we think our estimates are plausible.}. Symbolic computation has a very large impact through a variety of software packages and problem-solving environments.

The history of any important field is of intrinsic interest, and every researcher in it could benefit from knowing the main currents of its history.  Such knowledge also has instrumental value in that knowing the history can prevent new projects from repeating earlier mistakes and can help overcome similar obstacles. 
As the ironic saying goes, `Six months in the lab can save you three days in the library.'  

Finally, the history of symbolic computation is a crucial component of the history of mathematical thought more broadly since it was one way that ideas of abstraction and axiomatic thinking reached much larger audiences from the mid-20th century onward~\cite{Dick2020,steingart_axiomatics_2023}. 

This paper tells part of the story of the history of symbolic computation in a way that addresses these desires.  We have spent some time surveying, reading, and digesting the early literature on symbolic computation, and here we present some highlights of the period 1965--1975.  However, this paper is not comprehensive. We welcome discussion, especially of things we have missed. If you have knowledge that has passed us by please get in touch so that as full a story can be told as possible.

When SIGSAM was being founded those involved had to decide what to call
the new group. 
Sammet in~\cite{Sammet:PLHF:1969} says ``The phrase \textsl{formal algebraic manipulation} applies to the computer processing of formal mathematical expressions without any particular concern for their numeric values.''  Other writers wondered if formal logic should be included.
After discussion they ended up with ``Symbolic and
Algebraic Manipulation,'' well aware that some people would have used other
mixes of words. Similarly ISSAC ended up making ``Symbolic and Algebraic
Computation'' definitive at least for the first few decades. No group
wanted the choice of names they had made to be used to exclude
interesting work with relevance to their core interests, and gradually
the body of publications by each group defined by example what the field was
seen to encompass at the time.

By 1972, the field (whatever its name was) was beginning to be incorporated into the curriculum: 
see~\cite{CavinessCollinsSIGSAM72},
which was cross-published in the SIGCSE (Computer Science Education) Bulletin as 
well~\cite{CavinessCollinsSIGCSE72}.  
In there we find the field defined as follows:
\begin{quote}
\textsl{Symbolic Mathematics} is the study of algorithmic procedures for manipulating symbols that represent mathematical objects.
\end{quote}
We think most readers today would agree that this definition is not satisfactory, because (for instance) on its face it includes numerical mathematics. 

Things are a bit more clear-cut nowadays, and the name of the Journal of Symbolic Computation is meant to be inclusive, as discussed in detail in~\cite{buchberger1985symbolic}.
The \textit{Computer Algebra Handbook}~\cite{Grabmeier2002-yg} also tries to define what topics are relevant to computer algebra, in the more than one hundred and fifty pages and twenty sections of its Chapter 2, but even there some topics that \textsl{could} have been included---such as interval arithmetic and automatic differentiation---were not included, while some of the topics that \textsl{were} included might be controversial: \sysname{MathML}, for instance, is extremely useful for sharing mathematics, but is it ``Symbolic Computation''? 
It is still possible to disagree on whether a topic should be included or not. 

However, we do not attempt here to define what is, and what is not, a topic relevant to Computer Algebra even for the time period we consider, and instead we simply admit that we omit some topics in this paper. We used ``Symbolic Mathematical Computation'' in the title, to make it a bit clearer for outsiders.
At one end of the spectrum have been groups that have built software to be applied in a range of mathematical and engineering domains. At the other end have been those concerned with theory. To some extent, this can be seen as the continuation of earlier work to rebuild mathematics from
a constructivist standpoint. 
The mix of tensions and cross-pollination
across the different approaches is part of what gives the subject and its
associated conferences some of its flavour.

We have organized our report chronologically in sections, more-or-less ending in 1975 (about a half century ago). Activity in each period had its own distinctive dynamic. Within each section we use the device of classifying contributions as those for software systems, applications, algebra and formal proof.

By and large we \textsl{exclude} numerical computation here, except for multiple precision and except insofar as it was used for exact computation.  
See the impressive work~\cite{HistoryNLA:2022} for an 800-page journey through just the history of numerical linear algebra, as a partial recompense for this exclusion.

Our focus is on symbolic mathematical computation for programmable computers.  
We highlight key algorithmic developments in this context.
% ...

\begin{figure}
  \centering  \vspace{.25\baselineskip} \includegraphics[width=0.50\textwidth]{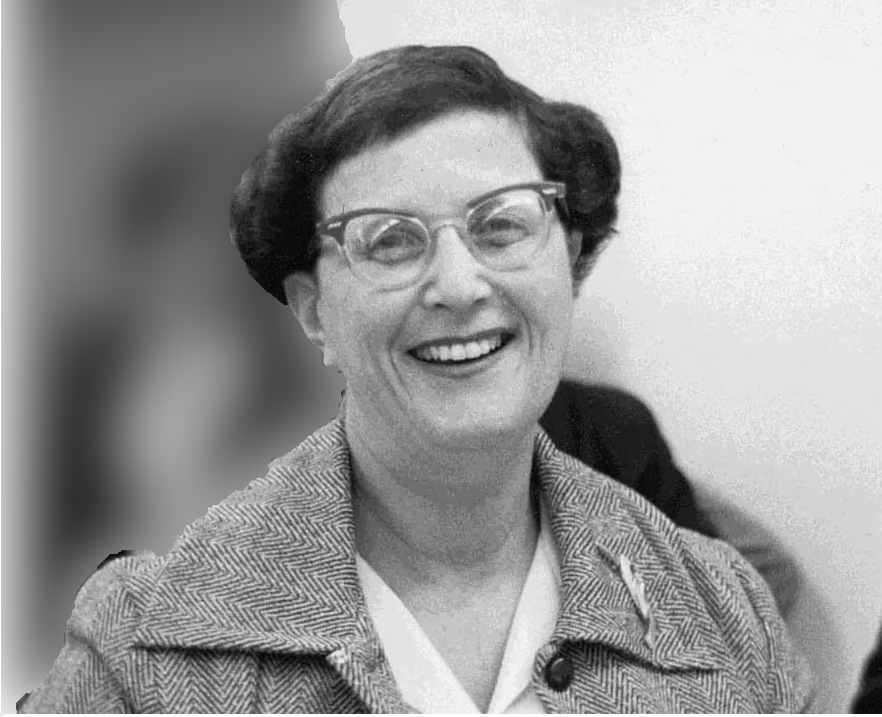} \vspace{-0.1\baselineskip}
  \caption{Jean E.~Sammet, the founding Chair of SIGSAM and General Chair and Program Chair of SYMSAC '66} \vspace{1.5\baselineskip}
  \label{fig:teaser}
\end{figure}

We have chosen to frame our work by focusing first on the contributions of Jean E.~Sammet (1928--2017), who contributed signally to the early symbolic computation world~\cite{Davis2024Sammet} and took prime responsibility for the SYMSAC'66 conference that was celebrated in Guanajuato as having grown to become ISSAC.  Among her further achievements, she was the founder and first Chair of SIGSAM, the ACM Special Interest Group on Symbolic and Algebraic Manipulation.  She was the lead designer of \sysname{Formac}, which was the first commercially successful programming language for computer algebra, and was instrumental in the development of \sysname{Cobol}. She later became the first female president\footnote{This is not so much indicative of Sammet's ability or force of personality, but instead an indictment of the times.  Some of the language of those times will necessarily creep into this paper: for instance, Sammet became the first ``Chairman'' of SICSAM, later SIGSAM.  This may read oddly to more modern sensibilities.} of the ACM itself. See figure~\ref{fig:teaser}.

Sammet was also a historian of computing. ``From childhood on, I hated to throw papers away. As I became an adult, this characteristic merged with my interest in computing history. As a result, I created important files and documents of my own, and became concerned with having other people publish material on their important work so the facts (rather than the myths) would be known publicly.''~\cite{fisher2017jean} She wrote an encyclopedic book~\cite{Sammet:PLHF:1969} on the early history of programming languages (concentrating on those invented in the USA), including those for computer algebra, and later wrote a condensed paper on the same subject, together with a look to the future~\cite{Sammet:ProgrammingLanguages:1972}. As well as her having started the SYMSAC
conference series and founding key early groupings and publications, her output has helped us to see how those active in the field saw it at the time, and in particular what they viewed as part of the subject: so what we report here is less tainted by our modern views than might otherwise have been the case.


For more details of her life, consult the \href{https://ethw.org/Oral-History:Jean_Sammet}{IEEE Oral History interview}\footnote{Jean Sammet, an oral history conducted August 2001 by Janet Abbate, IEEE \url{https://ethw.org/Oral-History:Jean_Sammet}. The ETHW has a collection of more than 900 oral histories in electrical and computer technology which can be accessed via \url{http://ethw.org/Oral-History:List_of_all_Oral_Histories}.} of her by Janet Abbate~\cite{Abbate2001}.  We learn that she never completed a PhD, leaving graduate school with only a Masters' in abstract algebra (from a pure mathematics department where computers were held in contempt---which is the word she used). She tried to become a math teacher, but met with insurmountable trivialities, and instead went into industry, first at Metropolitan, then Sperry Gyroscope, then Sylvania Electric, and finally IBM.  It was in industry where she learned to make wiring diagrams for punch-cards, which she found she really enjoyed, and to use analogue computers for military engineering problems, which she didn't much care for.  Eventually digital computers came into play.  The interview gives her explanation of how she taught herself to program, working at Sperry Gyroscope, and indeed taught some of the very first college courses in programming, before she took a position at IBM:

\begin{quote}
My boss’s boss, my manager, came over to me one day and said, ``By the way, do you know we’re building a digital computer?'' I said, ``I’ve heard of it. I’m not sure I know what that means.'' He said, ``We are building a digital computer, because we can see that that is the wave of the future.'' Because this was a company that made its living on contracts with the Federal Government, and in particular with the military, so they could see, in 1955, that that was going to be the wave of the future. He said, ``We’re building this thing. Do you want to be our programmer?'' I asked the obvious question: ``What is a programmer?'' And he said, ``I don’t know, but I know we need one!'' [laughs.] I looked at him, and I said, ``Well, is this anything like working with punch-cards?'' And he said, ``I’m not sure, but I think it might be.'' So I thought, ``Well, that punch-card stuff was fascinating. I’m not too inspired with this submarine stuff. I should at least give this a try.'' So I became the programmer. No books, no manuals, no instructions, no nothing---and engineers who somehow thought that machine was going to run itself and didn’t want a programmer. Fortunately, my boss, who was a very nice and very smart engineer, knew they needed one, even if he didn’t quite understand what that meant.
\end{quote}
The rest of the interview is well worth reading.

\section{1964 and before}
Of course there is a significant prehistory of computer algebra.  In~\cite{VELEZ2022} we find a nuanced discussion of the role of Charles Babbage himself as a potential ancestor, expressing himself in writing on the topic as early as 1836.  Ada Lovelace's later writings (1846) amplify Babbage's musings, and make it clear that the analytical engine was being viewed with a potential for algebraic, not just numeric, computation.  See~\cite{bowen1995brief} and~\cite{larcombe1999lovelace} for more discussion of this prehistory.

By the 1930s mathematicians had developed a concept of an
``effective procedure'' for completing a task, even though computers as we know them were not available. 
The 1926 paper of Grete Hermann~\cite{hermann1926frage}, translated to English in 1998~\cite{Hermann:Finitely:1998}, was cited in~\cite{BuchbergerCollinsLoos:1982} (the table of contents of which is in~\cite{Buchberger:BOOKREVIEW:1982}) as being of special importance. These century-old works really set up the
idea of what could be done algorithmically, and the biggest result associated
with it was G\"odel's Incompleteness Theorem which effectively dashed the
hopes of those seeking to rebuild all of mathematics on constructive foundations.
A key point to make here is that in deriving that result, G\"odel publicised the
fact that any formula could be encoded using a number. 

The Turing machine was
defined in the same decade. It established a framework for mechanical computation
that could handle mathematical equations and a desire to solve them
exactly or prove properties of them. It also suggested using one computational
model to simulate another: in effect introducing the concept of an ``interpreter''.
In the same time-frame, the American logician Alonzo Church introduced the lambda calculus as a model of computation.  
That the lambda calculus and Turing machines provide equivalent computational power is the Church-Turing thesis.
Church also proved that problem of deciding validity of formulas in first-order logic is unsolvable~\cite{Church:1936a,Church:1936b}, which has implications throughout computer algebra, including subsequent work on the decision problem~\cite{Davis1961} and later Richardson's theorem on the undecidability of symbolic real expressions~\cite{richardson1969some}.
Of course, all of that work was theoretical and much of it did
not see general algebraic transformations and simplifications as the main
objective, but it provided a basis for those who came later. 

By the late 1940s and through the first half of the 1950s an increasing number
of mathematicians became able to access electronic computers. Several of them 
had prior experience using electromechanical calculators
to compile tables of special functions or to predict the trajectories of
shells: work that was numerical in style. But some looked at problems that
might be described as pure rather than applied mathematics. Perhaps the main
focus was on support for logic, inference, and proof checking. Today these are still official topics of ISSAC, but in the early years 
schemes set up
to handle them became directly applicable to symbolic elementary algebra.

A notable contribution was \sysname{IPL} (Information Processing
Language) by Allan Newell, John Clifford Shaw, and Herbert A. Simon (c.~1956)~\footnote{See
the description of the contributions of Newell, Shaw, and Simon at the History of Information website. \url{https://www.historyofinformation.com/detail.php?id=742}},
which introduced list-based symbolic data structures, recursion, and dynamic memory
allocation~\cite{Sammet:PLHF:1969,Newell1958IPL}. 
\sysname{IPL} was developed in close connection with the
\sysname{Logic Theorist}, one of the earliest artificial intelligence programs, which demonstrated that machines could discover proofs of
theorems in symbolic logic~\cite{NewellSimon1956}.
In that system, logical expressions were represented as structured
symbolic objects with an explicit tree organization
(as described in the original report), and proof search was guided by heuristic strategies rather than exhaustive enumeration.
These requirements motivated the use of flexible symbolic representations
together with operations for dynamically constructing and traversing
such structures.
The \sysname{Logic Theorist} operated on formulas drawn
from \emph{Principia Mathematica} and was able to prove a substantial
subset of its early theorems~\cite{NewellSimon1956}.
Thus \sysname{IPL} can be seen not merely as a programming language
innovation, but as part of an integrated approach to symbolic
reasoning, in which representation and control mechanisms were
designed to support heuristic search in spaces of mathematical
expressions.
Newell and Simon demonstrated their automated reasoning program at 
the 1956 Dartmouth Summer Research Project on Artificial Intelligence, which was co-organized by John McCarthy.

\sysname{Fortran} made its first appearance in an IBM memo dated November 10, 1954, and \sysname{Fortran II} was released in June 1958.  See~\cite{Sammet:PLHF:1969} for a detailed discussion.
As will be seen in the next section, \sysname{Fortran} later
became an important base for mechanized symbolic computation. But from its
inception
(indeed from the time of \sysname{Autocode} some years earlier) 
the
issue of parsing algebraic formulae had been addressed. From technical reports
published even some years later it is clear that this was 
a challenging task for quite some time.  It is hard to have an algebra system without a way
for the system to accept reasonably natural mathematical input! Indeed some of the very earliest attempts used what now seem like bizarre tabular notations for both problem presentation and delivery of results.

\sysname{Algol} was developed in the period 1955--1957 by work started by a committee struck by GAMM\footnote{\url{https://www.gamm.org/}}~\cite[p.~173]{Sammet:PLHF:1969}.  \sysname{Cobol} was developed extremely rapidly, taking just the last six months of 1959~\cite[p.~331]{Sammet:PLHF:1969}. \sysname{Lisp}, very much a successor to \sysname{IPL}, originated at around the same time (the key publication~\cite{DBLP:journals/cacm/McCarthy60} was in 1960), and those working close to
McCarthy started to use it for a range of symbolic calculations. 


As with
\sysname{IPL}, some of the very earliest activity concentrated on logic rather than
algebra. Efficient list processing, the first \sysname{Lisp} compiler (1962), extended precision integers and floating 
point values (including elementary function evaluation), and garbage collection would eventually build on this to
provide necessary infrastructure for symbolic algebra.  

Somewhat later, the first in the \sysname{Snobol} family of string processing programming languages was initiated~\cite{farber+:1963,farber+:1964} at Bell Telephone Laboratories.  
This emerged from the work on \sysname{SCL}~\cite{lee+:1962,griswold:1978}, ``a Language for Symbolic Communication,'' used for symbolic integration, factoring of multivariate polynomials and the analysis of Markov Chains~\cite{griswold:1978}. 
Symbolic computation has continued to be one of the intended application areas of the \sysname{Snobol} languages.
For a summary of the early history of
programming languages, see~\cite{Sammet:ProgrammingLanguages:1972}, which has a beautiful map showing lines of descent.

The state of the art of symbolic computation software in the period covered by this section has been captured in Sammet's survey article~\cite{Sammet1966}.
On the algorithmic side, of particular note is the emergent interest in faster methods for multiplication by Karatsuba~\cite{KarOfm:1962} and leading to Toom-Cook multiplication~\cite{Cook:1966,Toom:1963}.

The situation at the end of the 1950s can be looked at in two quite
contrasting ways. Almost nothing that we would now view
unambiguously as an ISSAC topic had yet been published. But several projects
had begun and a very substantial body of fundamental work would lead to tremendous growth
the following decades. From that
perspective the bounty of the 1950s included:
\begin{enumerate}
\item Computers of increasing power and reliability became fairly broadly
accessible to people wishing to develop advanced applications;
\item High level languages rather than machine code became a practical
strategy for developers, with some hope for cross-platform portability.
\sysname{Fortran} and \sysname{Lisp} were the ones that were widely used then and have lasted, but
\sysname{Algol} existed and has influenced almost everything since, and at the time
string processing such as \sysname{Snobol} were seen as core technology for symbol manipulation;
\item List and tree data structures with automatic storage management was
an understood concept -- either built into the language used or implemented
as part of the program being written;
\item There were links between practical and theoretical work and workers relating to algorithms and computation. This spanned cost analysis of algorithms
where today the results would be reported in big-O notation, formal models
of computation such as the lambda calculus and forged bridges between theoretical development and practical problem-solving applications;
\item The body of experience in proof checking and calculations involving
logic was such that adapting it to work with general classical algebraic formulae could
seem natural;
\item Initial planning or work was under way on many significant projects that will be described in the next section. 
\item The concept of artificial intelligence inspired much work on carrying out
procedures that humans find challenging, and working with algebraic formulae certainly fell into that category.
\end{enumerate}

None of this work was isolated from wider social and cultural concerns. RAND Corporation, where
\sysname{IPL} was developed, had spun off from a post-war US Air Force project to plan for future weapons. We view a verification of Russell and Whitehead as perhaps an unusual future weapon and celebrate the support that project was given.
The Soviet launch of Sputnik in October 1957 led President Eisenhower to establish the Advanced Research
Projects Agency (ARPA) early the following year. One of its missions was to give the US the capacity to
launch and use spacecraft. This called for work on automating calculations in celestial mechanics and orbital dynamics where pure numeric computation became strained and symbolic work was the best option.
ARPA would go on to fund many of the projects we describe below.

\section{1965--1970}

At the instigation of Jean Sammet, the ACM Special Interest \textsl{Committee} on Symbolic and Algebraic Manipulation (SICSAM) was formed in 1965 by George Forsythe, then President of the Association for Computing Machinery. 
In a 2006 oral history~\cite{Bergin:SAMMET:2006}, Sammet recalled Forsythe writing back to her saying ``I think that’s a fine idea. You are now the Chairman...'' Years later, Sammet recalled that one of the main motivations to form SICSAM was to have a forum where staff from different companies could share ideas~\cite{Sammet:199x}. Later this ``Committee'' became a ``Group'' and ACM SIGSAM was born. The following year, Sammet organized the first SYMSAC, which took place in Washington, DC and eventually led to the ISSAC conference series. ``Everybody that went thought it was wonderful, and when you expected them to be out at the bar, they were in sessions!'' she also recalled.

Although the ACM Digital Library does not record who was the General Chair of that conference, Sammet said she served as General Chair, Program Chair, and Chair of SICSAM, ``a troika of one.'' Some of the papers were published in the 
\textit{Communications of the ACM} after the conference. They include works on systems, on applications, on techniques for differentiation, and something that at first glance \textsl{looks} like it's for teaching---it's entitled ``Grad Assistant''---but is not; the idea was that the program would replace your graduate student research assistant and do your drudge computations for you!

\noindent
One more quotation, this one with an ironic outcome.  Sammet said:
\begin{quote}
There was a very strong Special Interest Group on numerical analysis [SIGNUM] in ACM, and at some point I think I contacted them, and, in a presumably polite way, they told me to go away and not bother them. They were interested in numerical analysis, and they didn’t want any of this non-numerical stuff floating around. And that attitude lasted for a very, very long time.
\end{quote}
When SIGNUM closed up shop around the turn of the 21st century, during the tenure of one of the present authors as Chair of SIGSAM or shortly thereafter, its remaining assets were assigned to SIGSAM.  So the present ACM SIGSAM is the heir to both groups.

We mention one more item of interest from the archives of the SICSAM Bulletin (later the SIGSAM Bulletin, later \textsl{Communications in Computer Algebra}).  The first Editor of the Bulletin was Peter Wegner\footnote{Peter Wegner (1932--2017) was a professor of Computer Science at Brown University from 1969 to 1999. His work on object-oriented programming is considered seminal.}.  He passed the torch after a year to John Young, who announced in his first message to the readership\footnote{On p.~1 of Issue \#5}, that there would be a European sub-editor to whom correspondence should be addressed from that side of the Atlantic. That sub-editor was \href{https://en.wikipedia.org/wiki/Maurice_Wilkes}{Professor Sir Maurice Vincent Wilkes}, of Cambridge University.  Wilkes won the Turing Award the next year, in 1967, and was knighted in 2000 for his many contributions to computer science. Early issues of the Bulletin also included a note on curriculum design for computer science by Donald~E.~Knuth (ACM Turing award 1974) co-authored with Peter Wegner. We will also note that ``Tini'' Veltman, whose 1963 algebra system \sysname{Schoonschip}~\cite{Veltman1963} is another candidate for being ``first,'' went on to become a Nobel Laureate. It seems that Sammet was right: there was significant interest in the community in symbolic and algebraic manipulation, and that attracted participants of extraordinary calibre.

Sammet identified the language \sysname{Algy} as the first published step towards a general-purpose language for computer algebra~\cite{bernick1961algy}.  The authors were Myrna D.~Bernick, E.~D.~Callender, and J.~R.~Sanford.  The system \sysname{Algy} seemed to allow basic polynomial manipulation.  Sammet credited this work with ideas that led to the creation of \sysname{Formac} in this period.  \sysname{Formac} has been called the ``first commercially successful computer algebra system.'' Elaine R. Bond's history of \sysname{Formac}, published in the \textit{Proceedings of SYMSAC '66} acknowledged the influence of ``other similar systems such as \sysname{Alpak},
\sysname{Altran} and \sysname{Formula Algol}'' and cited a paper of Sammet's that appeared in technical report form in 1965, and later in~\cite{sammet1967formula}.

\sysname{Alpak} (Algebra Package) was a set of \sysname{Snobol} routines and macros for symbolic manipulation of very large rational algebraic expressions. It was implemented by W.~Stanley Brown and colleagues at Bell Labs in the early 1960s. This was succeeded by \sysname{Altran} ``a highly portable implementation of both algorithms and compiler... [which] included algorithmic advances in the handling of multivariate polynomials and macro generation to tailor \sysname{Fortran} code to make full use of the characteristics of particular hardware. Its features included a run-time environment with dynamic storage allocation, recursion, symbolic dumping, and error handling''. One challenge raised by the \sysname{Altran} team was finding multivariate greatest common divisors~\cite{Brown1966}. 
This led to intensive work in the early 1970s before being satisfactorily resolved~\cite{millman_history_1984}.

It's part of the folklore of the symbolic computing community that `the first program for symbolic differentiation was written before \sysname{Fortran} existed,' and this turns out to be true.  In~\cite{Bender:differentiate:66}, a paper describing a system ``\sysname{Manip}'' that could differentiate, we find citations that go back to 1962. That 1962 paper~\cite{Hanson:Analytic:62} cites two 1953 papers, one by Kharimanian and one by Nolan, and describes the process by which the programs had to be used: in each case the expressions had to be translated into a compact code before the programs could be run. These papers are credited in~\cite{Sammet:PLHF:1969} as being the first.

Drawing further from~\cite{Hanson:Analytic:62}, we infer several things.
First, these early differentiation programs were seriously limited by
the capacity of the computers used and a consequence was that the input
and output of expressions had to be in really ugly restricted formats.
The one in~\cite{Hanson:Analytic:62} accepted input in forms that looked like
\begin{verbatim}
    A * X P 2 -I- B * X -4- SIN.(C * X)
\end{verbatim}
They used ``P'' where today we might use ``\verb+^+''.
They parsed the input into a kind of {\L}ukasiewicz prefix form~\cite{lukasiewicz1931uwagi} in a table and
thought of this more as a table than as a tree.

The applications that they mention include a guided missile system, and they say ``Equations consisting of three hundred or more
symbols have been successfully differentiated with resultant derivatives of length in excess of seven hundred symbols. The equations under consideration involve six independent variables and four dependent variables.''

If one can differentiate, then one can compute Taylor series.  It turns out that for efficiency one should be careful, because naive methods lead to combinatorial growth in the length of symbolic expressions, as soon became painfully clear.  So-called \textsl{automatic} differentiation emerged later to address this, but early work included~\cite{reiter1965automatic}. That report described a program that generated \sysname{Fortran} and machine code for the particular computer in use at the author's institution in a way that we might classify as automatic differentiation today.

The papers so far highlighted here are concerned with system construction.  Papers from applications  also appeared in SYMSAC '66, such as~\cite{Cuthill:linear:66}, which described a \sysname{Formac} program to solve linear initial and boundary-value problems for ODE.  The author, Elizabeth Cuthill (1923--2011), is perhaps most famous for her work on the \textsl{Cuthill--McKee} algorithm, which permutes sparse matrices in a way to reduce bandwidth.  The paper describing that work is~\cite{Cuthill:Reducing:69}. It's a judgement call to say that that work was ``symbolic computation,'' but since it is so famous and so useful and has been cited thousands of times we are quite motivated to claim it for the community!

Another contribution in the applications line is~\cite{Danby:Poisson:66} which described manipulation of Poisson series, that is, series where the terms are of the form
\begin{equation}
    x_1^{k_1} x_2^{k_2}\cdots x_m^{k_m} e^{y_1}e^{y_2}\cdots e^{y_n}\>.
\end{equation}
Such series\footnote{Or the more or less equivalent ones with sines and cosines in place of the complex exponentials} were, and are, in great demand for solving problems in celestial mechanics; see the famous paper~\cite{deprit1969canonical}.  

James R. Slagle (1934-2023), who completed his 1961 dissertation under Marvin Minsky~\cite{slagle1961heuristic}, described a \sysname{Lisp} program called \sysname{Saint} which could find antiderivatives of many functions. Slagle, who was blind, was also a chess champion. He won the first Over-the-Board tournament of the US Braille Chess Association.\footnote{\url{https://web.archive.org/web/20160502160843/http://www.americanblindchess.org/potb.htm}} 

Joel Moses (1941--2022), who began his doctorate in 1963, redesigned Slagle's \sysname{Saint} into a symbolic integration program called \sysname{Sin}, completed in 1967. It was also written in \sysname{Lisp} and advanced the use of knowledge based systems rather than tree search for AI~\cite{moses2012macsyma}. This work in \sysname{Lisp} coincided with the creation of Project MAC at MIT (1963). Its first director was Robert M.~Fano. Out of this project the \sysname{Macsyma} system would be born.

\sysname{Mathlab} was also a product of the mid-1960s, created at MITRE\footnote{MITRE was a military think tank that was spun off from MIT Lincoln Labs in 1958. Its first employees were developers of the \sysname{Sage} system which provided air defense during the Cold War.} by Carl Engelman (1929-1983)~\cite{engelman1965mathlab}. One of his 
rules stated ``\sysname{Mathlab} is intended for the physicist, not the programmer.'', while another insisted:
\begin{quote}
The computer, as viewed by the user, must be
intimate and immediate. The user should have
next to [his] desk a console consisting of a typewriter
or, preferably, a typewriter and a scope.
Economy might, in some cases, dictate the substitution
of a plotter for the scope. These are connected
to a large, fast, on-line, time-shared
digital computer. [He] communicates with that
computer by typing messages on his typewriter or
by means of a light-pen on the scope. The computer
replies by means of the same machines. It
types both messages and equations. On the scope
it displays both equations and graphs. Above all,
the response time to the user's requests must be
short.
\end{quote}
It would be at least twenty to thirty years before that goal would be widely achieved. \sysname{Mathlab 2} introduced two-dimensional output (subscripts and superscripts in \sysname{Ascii} format).

\sysname{Scratchpad/1}, written in \sysname{Lisp} by James Griesmer, Dick Jenks and later David Yun~\cite{griesmer:71}, started in 1965 but was never publicly released.  First versions are important for history, though: Griesmer's knowledge was later incorporated into \sysname{Scratchpad II} by Dick Jenks and his team.

\subsection{The gospel according to Moses}
In about 2010, Joel Moses published his memoirs, entitled ``My Life.''  These memoirs, which cover a much wider time period than we are restricting ourselves to in this paper, seem to have been published only online, at \url{https://www.csail.mit.edu/sites/default/files/2022-10/moses_memoirs.pdf}, and seem not to be widely cited (at least according to Google Scholar).  We found these memoirs to be a significant source of information, particularly about the growth and influence of computer algebra in the United States but also about American academic culture of the time.  We highly recommend reading them yourself, and we restrict consideration here to only a few points among many possible.  We have entitled this section of our paper mostly as a friendly echo of Moses' sense of humour, which you may find examples of in his memoirs, but also as a reminder that in this section we are reporting his views, not necessarily our own.

Joel Moses obtained his PhD under the supervision of Marvin Minsky and Seymour Papert\footnote{This present paper does not pursue the extremely important issue of the use of computer algebra or symbolic computation in education.  Any paper that did so seriously would have to spend significant time on the work of Seymour Papert.}, and subsequently had several PhD students (sometimes co-supervised) who themselves had significant effects on our field and that of AI more generally, including Jamil Baddoura, Richard Fateman, Michael Genesereth, Barry Trager, Paul Wang, and David Yun.
Moses published several influential papers, including~\cite{moses1971symbolic,moses1972towards} and~\cite{moses1971algebraic}, ``Algebraic simplification: A guide for the perplexed,'' which is possibly his most highly-cited paper.  


Moses' classical and religious training coloured his views on artificial intelligence, in a way that may still be relevant today.  He believed in the value of what he called ``structured knowledge'', meaning knowledge structured in layers, and he made a distinction between this and knowledge structured by use of hierarchical trees.
His documentation of this belief, in his memoirs, includes tracing his family lineage back to the sixteenth century Rabbi of Prague, Rabbi L\"owe, also called the Maharal, who is reputed to have created the Golem.  Moses says that several of the researchers at MIT involved in AI in the early days also had been told family stories that they, too, were descended from this Rabbi, and that this had formed part of their motivation to study artificial intelligence.  Indeed, somewhat whimsically, he dedicated his 1967 PhD thesis ``To the descendants of the Maharal
who are endeavoring to build a Golem.''

On an even more whimsical note, we remark that the highly-cited simplification paper mentioned above was titled in clear echo of the 12th century work of Moses Maimonides, ``The Guide for the Perplexed.'' 

Moses claims that he was the one who chose the name \sysname{Macsyma}:
\begin{quote}
In the fall of 1968 Bill Martin, Carl Engelman and I began work on a project to
combine our three prior efforts to create a mathematical system that would be the
best of its kind. Bill led the effort to design the front-end and overall architecture
of the new system. [...]
I came up with the name of
the new system, Macsyma. Macsyma stood for Project MAC’s SYmbolic
MAnipulator. It also had an obvious Latin meaning, and a Hebrew one related to
the Persian word ``kismet'' and meaning ``magical'' and ``wondrous.''
\end{quote}
Phonetically, ``\sysname{Macsyma}'' invokes the Latin ``maxima,'' which is neuter plural substantive for ``the greatest things''.
Google Translate simply says that ``macsyma'' in Latin means ``maximum.''  It does not, apparently, know that it means magical or wondrous in Hebrew, although it does suggest ``meksima'' which it translates as ``charming.''  Perhaps Google is perplexed.



Early in his career, Moses met Jean Sammet: indeed, she hired him to work on Project Mac.  He had this to say about her:
\begin{quote}
    ``Jean was a determined person. She would see the project to completion,
and use it as a stepping-stone to increasingly greater roles in the computer
society, ACM, the Association for Computing Machinery. She eventually became
President of the ACM and a member of the National Academy of Engineering.
One of her major achievements, in my view, was that she nearly single-handedly
created the group of people who were interested in symbolic algorithms and
systems, SIGSAM, the Special Interest Group for Symbolic and Algebraic
Manipulation, becoming its first chairman. I became its third chairman a few
years thereafter. IBM lent its considerable stature to the nascent field of symbolic
computing with its support of \sysname{Formac}. Unfortunately \sysname{Formac} was a relatively poor
product, but that gave the rest of the community of system builders a
considerable advantage.''
\end{quote}

Moses volunteered to organize the second conference in computer algebra, in 1971, five years after the first one.

\begin{quote}
    ``The symbolic algebra conference took place in July of 1971, with about two
hundred in attendance--a good turnout for our field. I hurried from one session to
another to make sure things were going well, rarely staying for a full session or
even a full invited talk. I did make an exception for the talk on factorization of
polynomials by Elwyn Berlekamp of Berkeley, who was something of a
showman. He made it appear that he did not know what he was doing, and all
the while he pulled technical rabbits out of the air. I loved his talk. In fact, the
invited talks overall were of very high quality, as were most of the submitted
papers.''
\end{quote}

Moses continued to work on integration, and on simplification, and on special functions in symbolic computation, throughout his non-administrative career.  On the administrative side he rose to be Provost of MIT, and many pages of his memoirs deal with that, throwing around budgetary numbers that sound significant even today.

We draw attention to the following passage, on p.~221 in the section on the ``Athena project'' which was (apparently) the first serious attempt to use symbolic computation in education\footnote{As we have stated, we are not pursuing education in this paper; we hope to document this aspect in future work.}:

\begin{quote}
    ``Gerry also did not want us to do significant system development for
Athena. I had hoped that \qsysname{UNIX} would fill our needs, but I overestimated its
capability in a large and complex environment such as Athena. We were saved
by Jerry Saltzer, who became the technical director of the project. Under Jerry’s
supervision we introduced many improvements to \qsysname{UNIX} that have become
standards in the \qsysname{UNIX} world. One was Kerberos, the software for authenticating
users in the network. A key related addition was \sysname{X-Windows}, initiated by a
graduate student in \sysname{LCS}, and championed by a staff member from DEC, and
approved by me for continuing development by Athena. \sysname{X-Windows} permitted
\qsysname{UNIX} systems to emulate the windowing capability of a MAC computer. DEC felt
that \sysname{X-Windows}, which became DEC Windows in their environment, justified their
entire investment in Athena.''
\end{quote}
One of the points of this present paper is to document places where computer algebra or symbolic computation has had an impact beyond the field.  This passage reflects two important such instances.

\begin{figure}
    \centering
    \includegraphics[width=0.45\linewidth]{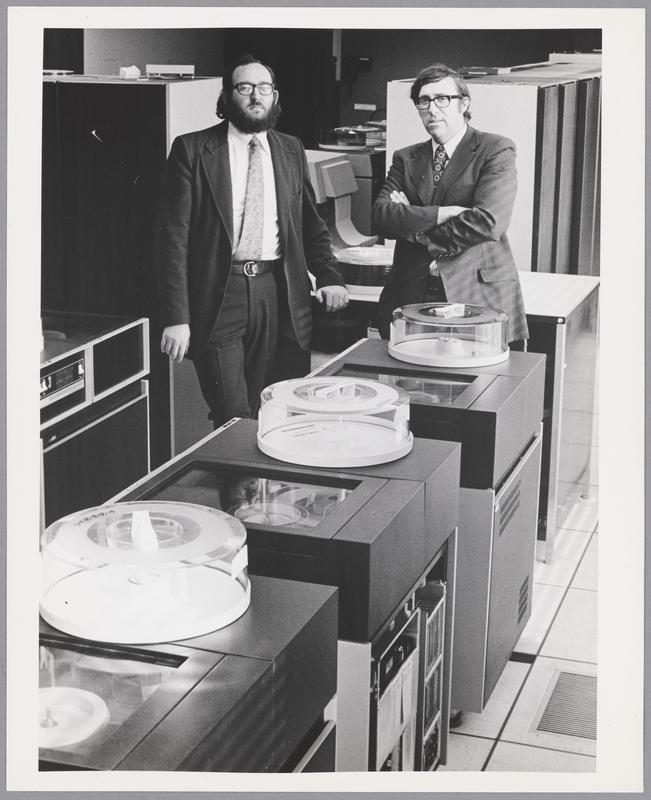}
    \caption{Joel Moses (left) and Michael Dertouzos (right) in the Laboratory for Computing at MIT in 1974.  Photo (c) MIT 2010, courtesy MIT Museum.}
    \label{fig:MosesDertouzos}
\end{figure}


There is a lot more to think about, as written in those memoirs.  Moses did summarize some of this material for a later JSC paper, namely~\cite{moses2012macsyma}, but we believe that the richer memoirs are well worth reading, and we leave off here.  You can also find an interview with him at \href{https://infinite.mit.edu/video/joel-moses-phd-%E2%80%9967/}{the MIT infinite video archive}.


\subsection{The world outside MIT}


\begin{figure}
    \centering
    \subfigure[Hearn]{\includegraphics[width=0.35\linewidth]{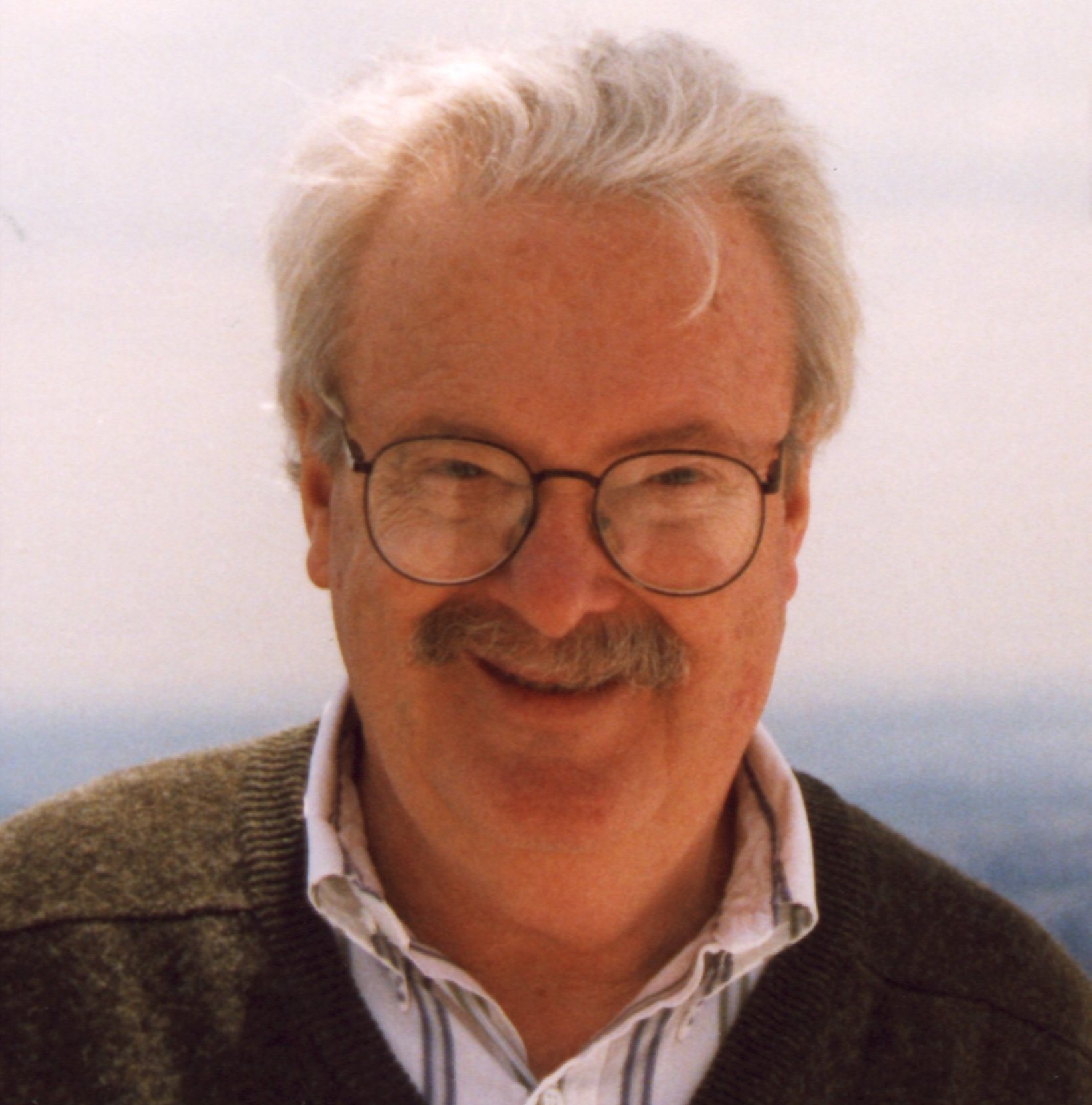}}
    \subfigure[Goto]{\includegraphics[width=0.5\linewidth]{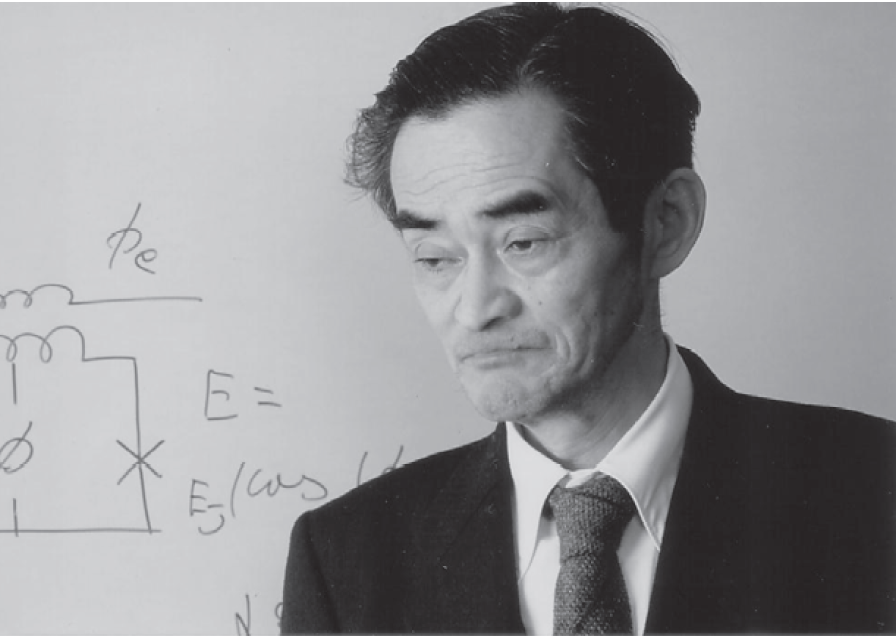}}
    \caption{[left] Anthony C.~Hearn. Photo by A.~C.~Norman. [right] Eiichi Goto.  Photo taken from~\cite{Ida2005}.}
    \label{fig:Hearn}
\end{figure}

\sysname{Reduce} was created in the early-mid 1960s~\cite{Hearn:68,Hearn:05} by Anthony C.~Hearn.  See Figure~\ref{fig:Hearn}. The first published \textsl{use} of \sysname{Reduce} was~\cite{Hearn:66}, where it was reported that six months of human labour had been reduced to fifteen minutes of computer time.  \sysname{Reduce 2} appeared in 1970.  One of 
the present authors
has the full working source of that program, and its manual, from which we find that \sysname{Reduce 2} did not have arbitrary-precision integers or bigfloats, and instead relied on whatever floating-point system was supplied by the underlying machine.  Given the state of floating-point arithmetic in the years before the IEEE Standards, this seems natural. \sysname{Reduce} had to wait until 1979 for a bigfloat package~\cite{DBLP:conf/eurosam/Sasaki79}. \sysname{Cobol}, in contrast, already had facilities for specifying the number of figures before and after the decimal point fairly freely, which was another milestone for Jean Sammet. The first (and probably only) algebra system implemented in \sysname{Cobol} was described in 1976~\cite{fitch1976design}. To be fair, that system, while slow, actually had some quite interesting features.

These were not the only systems being contemplated at this time.  John Cannon wrote a critique of algebraic programming languages in 1969, namely~\cite{Cannon1969critique}. Cannon wrote that his ``experience in developing programs to work with commutative and non-commutative polynomial rings'' informed that critique, and doubtless formed a foundation for the development of \sysname{Cayley}, released in 1982 (and thus outside the scope of this paper) and later \sysname{Magma}. Then there was the language \sysname{Formal} (a ``formula manipulation language'') developed for NASA/JPL prior to 1970 by Charles Mesztenyi~\cite{mesztenyi1970formal}, who later went on to help develop \sysname{GRAAL}, a graph processing language.

\begin{figure}
    \centering
    \includegraphics[width=0.5\linewidth]{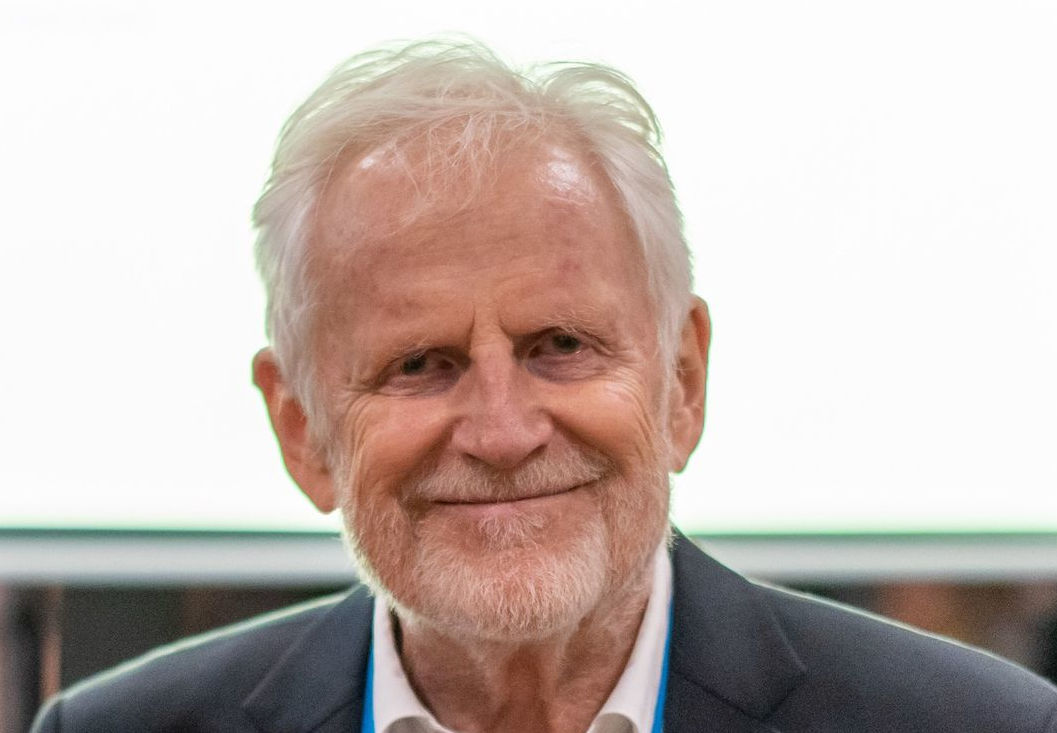}
    \caption{Bruno Buchberger.  Photo courtesy Tom\'as Recio, taken on the occasion of Bruno's 80th birthday. }
    \label{fig:Buchberger}
\end{figure}

%
Now, we must include a seminal contribution in the area of algebraic algorithms,
namely the 1965 PhD thesis of Bruno Buchberger~\cite{buchberger1965algorithmus} which introduced Gr\"obner bases to computer algebra, and which was later translated to English in~\cite{BUCHBERGER2006Translation}. See Figure~\ref{fig:Buchberger}.  The thesis was first followed by~\cite{Buchberger1970}, which itself was translated to English in~\cite{buchberger1998Groebner}. These major works and the Buchberger algorithm did not make their way into computer algebra systems until much later; in fact, not until after 1976 and the publication of~\cite{Buchberger:1976}.  Nevertheless these were important milestones for the era.

Before Gr\"obner bases, though, there was elimination theory using resultants.  Significant prior work includes~\cite{Habicht1948}. One very important paper of the late 1960s, now less well known than it deserves to be, is~\cite{Ku:Resultant:1969}. It made the claim that, for multivariate polynomials, B\'ezout matrices are superior to subresultants both in respect to speed and in respect to numerical stability.  In view of the very pessimistic recent result~\cite{noferini2016numerical} this may merit a second look with a modern lens. 

Other important papers in polynomial arithmetic 
include Berlekamp's factorization of polynomials over finite fields~\cite{berlekamp:1967} and Lipson's use of Chinese remaindering~\cite{lipson1971chinese} for homomorphic methods.

A fundamental and practical advance was made in~\cite{edmonds1967systems} and independently in~\cite{bareiss1968sylvester}, where an efficient method for solving linear systems of equations over the integers was given.  This was the first of the ``fraction-free'' methods. Later, in~\cite{bareiss1972computational}, this was extended to general integral domains.  The first paper, by Jack Edmonds, which already gives its result for general integral domains, apparently gives a counterexample to what is known as ``The Matthew Effect'' wherein the more famous scientist is given most of the credit; here, although Bareiss is arguably less famous than Edmonds (who, for instance, won the inaugural von Neumann Theory Prize in 1985), the algorithm is widely known as Bareiss' algorithm.  This seems to be true in spite of the fact that Edmonds' paper has 512 citations in Google Scholar at this time of writing, nearly the same as Bareiss' paper which has 538.  Bareiss did publish an internal report on this in May 1966 for Argonne National Lab~\cite{bareiss1966multistep}, while Edmonds' published work is dated November 1966 on the paper itself, although the publication date of the journal seems to be 1967.  However, according to the Matthew effect (a play on the Gospel theme from Matthew 13:12 ``For whosoever hath, to him shall be given, and he shall have more abundance: but whosoever hath not, from him shall be taken away even that he hath.''), Edmonds should have been able to acquire the lion's share of the credit anyway.  This, of course, is not to be confused with ``Stigler's Law'' which says that the name given to any concept is never that of the scientist who originally thought of it.  Michael Monagan of Simon Fraser University informs us that he is now calling it Bareiss/Edmonds fraction-free Gaussian elimination. \href{https://en.wikipedia.org/wiki/Bareiss_algorithm}{The Wikipedia page for the ``Bareiss Algorithm,''} as it is currently known, gives a link for a crisp scan of the book~\cite{bareiss1966multistep}, which might otherwise be hard to find.

A huge development in the theory of integration in finite terms was published in this time period, namely the papers of Robert H. Risch giving (in outline) an algorithm 
to either find an antiderivative of a given elementary function, or to prove that no such expression was possible. 
See~\cite{risch1969problem},~\cite{risch1970problem}, and~\cite{rosenlicht1972integration} for a more accessible introduction.  See also~\cite{rosenlicht1968liouville}, which seems to have been the paper which converted the earlier analytic techniques of Liouville and of Ritt into the algebraic terms that we know now.
Implementation of this algorithm took quite some time and effort; the \textit{Proceedings of EUROSAM '79} contain four important papers by Norman, Davenport, Trager, Moses and Zippel which helped to bring the development to a more satisfactory state.  Of course, research continues today.


At the risk of venturing too far outside of the symbolic computation world, we mention the pioneering work of Stephen A.~Cook on the complexity of computation of functions, beginning with multiplication, for example the 1969 paper~\cite{Cook1969}.  
This was soon followed by fast multiplication by Sch{\"o}nhage and Strassen~\cite{schonhage1971fast}.
The paper by Strassen~\cite{StrassenMult:1969} was a fundamental advance in improved complexity for linear algebra.
The field of ``computational complexity'' has had a profound effect on research in computer algebra ever since.  See, for instance, the textbook~\cite{Gathen:MCA:2003}.

We now mention the 1969 undecidability result of Richardson~\cite{richardson1969some}, and the 1970 paper of Caviness~\cite{caviness1970canonical}.  According to Joel Moses~\cite{moses2012macsyma}, Richardson's result was a bit controversial at the time because of his inclusion of ``absolute value'' in the collection of functions allowed in the expression to be processed.  Nonetheless, the fact that recognizing zero is undecidable over such a simple class of expressions remains remarkable. Specifically, if the expression $E$ contains $\ln 2$, $\pi$, $\exp(x)$, and $\sin(x)$, then the problem of determining if $E < 0$ is undecidable.  Adding the function $|x|$ to the list of possibilities makes determining if $E=0$ undecidable.  
As discussed in~\cite{caviness1970canonical} this result has many implications for practical symbolic computation, because simplification is fundamental. Because (some) simplification problems are undecidable, we must sometimes rely on heuristics, and this leads to a forest of difficulties even today.

\section{1971--1975}

The very efficient and compact \sysname{Camal} (for Cambridge Algebra Language), by David Barton, Stephen Bourne\footnote{Stephen R.~Bourne is known for the Bourne shell, for work on \sysname{Algol68}, and for having been President of the ACM. See his Wikipedia page \url{https://en.wikipedia.org/wiki/Stephen_R._Bourne}.}, John Fitch and others, started its gestation in about 1968 according to the delightful little history~\cite{Fitch:CAMAL40:2009} (see also~\cite{Bartion:scheme:1967,Barton:simple:1968}) but was not published until 1971.

\sysname{Macsyma} became publicly available in 1971, after developments mentioned in the previous section.  The history of \sysname{Macsyma} is sketched broadly in~\cite{Dick2020}, from the point of view of a historian of computing.  The comments there might be fascinating to the insider: we see that in addition to providing tools for human use, \sysname{Macsyma} forced the humans to adapt to the tools. HAKMEM (a February 1972 AI Lab memo) also came out of MIT around this time. Compiled by Michael Beeler, R.~William Gosper, and Richard C.~Schroeppel, it described `little known data' of interest to computer hackers, `to save some duplication of effort--except for fun'~\cite{HAKMEM}. It has become known to subsequent generations through references like~\cite{levy_hackers_1984} and~\cite{warren_hackers_2002}. \sysname{Macsyma} was weighted more heavily towards the knowledge of special functions than were the other systems at the time, and this memo reflects an important emphasis of that software system.  

\sysname{Altran} continued its development during this time.  We note that Morven Gentleman (1942--2018) is credited in the fourth edition of the \sysname{Altran} manual as being one of the developers, probably in the early 1970s; some of us knew him later as an advisor for the creation of the 1980s program \sysname{Maple}. Gentleman published a paper in the SIGSAM Bulletin using \sysname{Altran} for Truncated Power Series~\cite{DBLP:journals/cca/Gentleman74}.
Another related result was Gentleman showing that, unexpectedly at the time, the optimal multiplication
chain for powering a polynomial was repeated multiplication by the original polynomial~\cite{gentleman:1972mult}.

Also in this period, George Collins (1928--2017) introduced his \sysname{SAC-1} system for polynomial arithmetic~\cite{Collins:SAC1:1971}. This system, written in \sysname{Fortran}, 
included functions for polynomial GCD, factorization, resultants, exact real zero calculation, partial fraction decomposition, rational function integration, and solution of systems of linear equations with polynomial coefficients.

\subsection{Tony Hearn and \sysname{Reduce}}
There was a serious change in Cambridge brought about through the agency
of Tony Hearn. The ARPAnet was very young and a transatlantic link joined
up to London. In Cambridge and with the excuse of a new multi-way
collaboration that Hearn was setting up (him in Utah, but Goto's group in
Tokyo, the Cambridge few and David Stoutemyer in Hawaii) 
Cambridge got a 110 baud
teletype with a modem which and could dial the London node and connect
across to the Utah machine.

Quoting from \url{https://www.sciencemuseum.org.uk}
under objects-and-stories and arpanet-internet\#what-was-the-arpanet:
\begin{quote}
``In the UK, the National Physical Laboratory (NPL) created its own packet switching 
Local Area Network in 1971.

Then, on 25 July 1973, a computer at University College London (UCL) sent packets of
digital information to another in California. Led by Peter Kirstein at the Institute 
for Computer Science, UCL, this was \sysname{ARPAnet}’s first transatlantic link. 

It ran from London to Norway via cable, then on to the Seismic Data Analysis Centre 
(SDAC) in Virginia via satellite link, and finally across the US network to the 
Information Sciences Institute at the University of Southern California.

The \sysname{ARPAnet} team had originally identified NPL as its first London link, but in the 
context of developing relationships within Europe it was deemed politically 
unacceptable for a British government research laboratory to connect directly 
with the US Department of Defense.''
\end{quote}

\noindent
At that time the \sysname{ARPAnet} had
around 2 dozen nodes.

Involvement in that project helped justify grant funding at Cambridge that got Mary-Ann Moore
and Arthur Norman  visiting David Stoutemyer in Hawaii to work on 2D printing on a fairly dumb terminal for the algebra code he had
at the time\footnote{This 2D printing also had an elegant line-breaking algorithm which made use of labels for subexpressions.}, as well as a multivariate polynomial GCD based on a subtractive remainder sequence. Norman went on to Tokyo which was when he first met
Kanada, Sasaki, and other members of Goto's group. Subsequently, several Japanese
researchers arranged to visit Cambridge for various non-trivial amounts of
time. 

Tony Hearn visited Cambridge for a summer at a stage that John Fitch and Arthur Norman were
just getting \sysname{Cambridge Lisp} working well enough to be able to try to mount
\sysname{Reduce} on it. 
Arthur Norman recalls an anecdote from that episode, which is that an attempt crashed and after
some digging it turned out that a function that had been compiled into
machine code had just been redefined. A consequence was that the code
involved happened to be being executed but the reference that kept it
alive was replaced with a pointer to the new version---and his code garbage
collected compiled code chunks as well as simple data. So the only
reference to this code was a return address into it on the call stack.
When the return happened, the code had gone, and the memory was used for
other things. So he needed to adjust the garbage collector so that it found return addresses
and tagged code blocks they pointed into as `alive'.

\sysname{Reduce} users were early adopters of international networking, and that
over time this grew so that \sysname{Reduce} had contributions from many parts of
the world from China and Japan to the West of Utah right out to Russian
scientists to the East, with Tony Hearn running a sort of shuttle
diplomacy visiting sites where he could assist both physicists working out
how to solve their problems and potential or actual contributors of
packages that added capabilities. The portability of his code and the very
open way in which it was circulated built a community that is perhaps best
seen as foreshadowing the open source models commonplace today. This community
was very much not centralised at one particular institution.

This growth of community was somewhat different to that of \sysname{Macsyma} which was more centralised at
project MAC and probably mostly made available by remote access to the
project MAC computers\footnote{There is some discussion of this in Moses' memoirs, who mentions the early role of Richard Fateman at Berkeley in encouraging dispersal of \sysname{Macsyma}.}. In contrast \sysname{Reduce} was, already at the time, running on multiple different and
sometimes eccentric computers---indeed FLATS was purposely being built with \sysname{Reduce} in
mind---and this was happening globally, in groups each of which was individually probably
fairly small.

\subsection{Scratchpad/1}
It is hard to overestimate the influence of \sysname{Scratchpad II} project  in the 1980s that followed \sysname{Scratchpad/1} from the 1970s. The heady ``pure research'' environment of IBM T.~J.~Watson Research Center provided a hub for the exchange of ideas and technology. The interchanges between visiting researchers and all kinds of scientists and engineers led to many downstream effects.  The principal feature that emerged from \sysname{Scratchpad II} and that differentiated it from other computer algebra systems (before or since) was its formalization of the domains of computation.  This enabled both rigour and efficiency in a way that was both challenging (to the user, sometimes) and satisfying to the researcher.  

An example paper from the 1970s is~\cite{norman1975computing}. This paper appeared in ACM \textit{Transactions on Mathematical Software} (TOMS), which even then was an excellent vehicle for publication\footnote{Some of the present authors feel that this journal was not as well-used historically by the symbolic computation community as it might have been, perhaps because it was viewed as primarily a ``numerical'' software journal. This (very symbolic) paper shows that that perception was incorrect. There were a few further symbolic computation papers in TOMS since this paper, and they tended to have good impact outside our community, but there were not that many. More might have been better.}.  A comparison to the report~\cite{reiter1965automatic} published ten years earlier shows considerable development: we see for instance what is now called ``lazy evaluation,'' as well as considerably improved ease of use.  Portability, however, remained an issue for both works, although for completely different reasons. 

\subsection{Algebraic algorithms}
Leaving systems for the moment and considering algorithms, major progress was made in this time period on the important problems of multivariate GCD~\cite{moses-yun:1973} and factorization~\cite{Wang_1973,wang1975factoring} using ideal-adic methods~\cite{yun1974hensel}.
This work---building on the work of others such as Hans Zassenhaus~\cite{zassenhaus1969hensel} as it does---marks the take-off of sophisticated modular methods in computer algebra systems, which remain crucial system tools for efficiency.  
This was followed up by an influential Mathematics of Computation paper by Paul Wang and Linda Preiss Rothschild~\cite{wang1975factoring}, again showing increasing impact of symbolic computing on the wider community.

Several people have since observed that, \textsl{in applications}, multivariate polynomials factor frequently and usefully. This implies that the results of these fundamental papers have had, in all likelihood, an outstanding if underappreciated effect on the utility of symbolic computation.  Since multivariate polynomials with coefficients ``chosen at random'' mostly do not factor at all, perhaps the extreme utility of this work could not have been anticipated.

\subsection{The beginning of theorem-provers}
Perhaps the most highly-cited work ever published in the SIGSAM Bulletin was George E.~Collins' follow-up to~\cite{collins_quantifier_1974}, namely the abstract~\cite{collins1976quantifier} which is just outside our time frame, although the first one is inside our time frame and is the one that has the actual results (the second is just an abstract\footnote{At times one gets the idea that citation culture is irrational. We use Google Scholar to count citations because of its wide disciplinary coverage. Although flawed, Google Scholar citation counts are just a proxy for impact anyway, and because we use this consistently at least it's a reasonably fair comparator. Being off by, say, ten percent is entirely likely, but with this many citations that doesn't matter.}).  As of January 2025 that abstract had been cited more than two and a half thousand times.  This reflects the importance of the topics of cylindrical algebraic decomposition and quantifier elimination. Collins cites Tarski~\cite{tarski1948decision} in that first-mentioned paper for the foundational work.

\subsection{Continued growth of applications}
Descriptions of other applications of computer algebra continued to be published in this time period.  For instance, consider~\cite{barton1972applications}, a very substantial paper of over seventy pages (including a seven-page bibliography). This paper surveyed various applications in physics: celestial mechanics, general relativity, and quantum electrodynamics.  It also gave an overview of computer algebra systems for physicists (not just \sysname{Reduce}), and from this paper\footnote{We might have remembered this from~\cite{moses1971symbolic}, of course.} we learn that \sysname{Sin} by Joel Moses (mentioned above) already used Hermite reduction to perform integration by partial fractions.

\section{Towards more modern times}
We leave off discussion of the history of symbolic computation with a few short remarks leading to the present day.  We do not wish to give the impression that we believe that the current state is uniformly better than that of the past.  Indeed we do \textsl{not} hold that opinion, and instead lament some lost opportunities (for efficiency, for example).  But there is no question that the fifty years since 1975 have wrought great change and in many cases very significant advances.  We do not tell that story here even in the laconic style we used for the years 1965--1975, but rather just mention a few items.
\begin{itemize}
\item The first is the 1982 book edited by Buchberger, Collins, Loos, and Albrecht.  The table of contents is printed in~\cite{Buchberger:BOOKREVIEW:1982}, where we see that a comprehensive overview of the field is attempted for the first time. A significant collection of historical material is present in the volume itself. This clearly marks a degree of maturity in the field;
\item The first textbooks on computer algebra (it may be invidious to mention any particular one, but for instance the book~\cite{Davenport:88a} was translated to many languages);
    \item Wen-ts\"un Wu\footnote{In the Western manner of family name last, his name is spelled variously Wenjun Wu, Wen-Tsun Wu, Wen-tsun Wu, and Wen-ts\"un Wu.  He himself used the latter, and so we do the same.} and the method of characteristic sets 1978~\cite{Gao:WTWu:2017}, starting the topic of automated deduction in geometry by computer algebra tools that is still quite active (ADG conference series, see \url{https://adg-foundation.info}, \url{https://adg-foundation.info/conferences.html});
    \item The founding of the \textit{Journal of Symbolic Computation} in 1985;
    \item The founding of the Research Institute for Symbolic Computation (RISC)-Linz in 1987, of the Key Laboratory of Mathematics-Mechanization (KLMM)-Beijing, also in 1987, and of the Ontario Research Centre for Computer Algebra (ORCCA)-London, Ontario, in 1997;
    \item The rise of the most popular modern commercial systems, namely \sysname{Maple} and \sysname{Mathematica}, and the \sysname{Matlab Symbolic Toolbox} (initially using \sysname{Maple}).
    \item The creation and development of \sysname{Cayley}, later \sysname{Magma}, \sysname{Macaulay}, \sysname{CoCoA} and \sysname{Singular}. These last three were all started in the 1980s and specialized in computational algebraic geometry;
    \item The D5 principle, as expounded by Dominique Duval and others~\cite{dicrescenzo1988algebraic,della1985new};
    \item The merging of the SYMSAC conference series%
    \iflongbib~\cite{ SYMSAC1966,SYMSAC1971,SYMSAC1976,
    SYMSAC1981,SYMSAC1986}\fi~
    (held every 5 years from 1966 to 1986) with the interleaved European EUROSAM/EUROCAM/EUROCAL stream%
    \iflongbib~\cite{
    EUROSAM1974,EUROSAM1979,
    EUROCAM1982,EUROCAL1983,EUROSAM1984,
    EUROCAL1985v1,EUROCAL1985v2,
    EUROCAL1987}\fi~
    to form the named ISSAC stream in 
    1988\iflongbib~\cite{
    ISSAC1988, ISSAC1989,
    ISSAC1990, ISSAC1991, ISSAC1992, ISSAC1993, ISSAC1994,
    ISSAC1995, ISSAC1996, ISSAC1997, ISSAC1998, ISSAC1999,
    ISSAC2000, ISSAC2001, ISSAC2002, ISSAC2003, ISSAC2004,
    ISSAC2005, ISSAC2006, ISSAC2007, ISSAC2008, ISSAC2009,
    ISSAC2010, ISSAC2011, ISSAC2012, ISSAC2013, ISSAC2014,
    ISSAC2015, ISSAC2016, ISSAC2017, ISSAC2018, ISSAC2019,
    ISSAC2020, ISSAC2021, ISSAC2022, ISSAC2023, ISSAC2024, ISSAC2025}\fi
    , together annual since 1981 (for conference chairs and program chairs see~\cite{ISSAC-list});
    \item Developing visions of the future~\cite{boyle1988future,Watt-Future:2009};
    \item \sysname{Macsyma}, \sysname{Axiom} and \sysname{Reduce}, all of which had previously been either
    commercial or otherwise subject to redistribution constraints, were 
    re-licensed as open source, and the general purpose capabilities they offered
    thus became available to all cost-free. This also means that the algorithms
    and techniques embedded within them could be studied by anybody who was
    interested, and various volunteers support them right up to the current day. These will now count as among the oldest software products still being maintained and used;
    \item Continuing development of applications (for instance, the \sysname{Canadarm}\footnote{\url{https://www.asc-csa.gc.ca/eng/canadarm/}, \url{http://www.maplesoft.com/view.aspx?SF=141144/ChangingFaceRobotic.pdf}})
    \item Continuing development of practical methods for exact solution of linear systems~\cite{dixon1982exact};
    \item Portmanteau mathematical, engineering and educational support software, led by \sysname{SageMath} where symbolic computation is supported alongside numerical work, visualization and document preparation and presentation; 
    \item Hardware design motivated by the special needs of our subject, including both the Xerox D-series machines, the \sysname{MIT Lisp} machines\footnote{See \url{https://en.wikipedia.org/wiki/Lisp_machine}. See also the discussion at \url{https://en.wikipedia.org/wiki/Symbolics} and the adversarial connection to the modern GNU project.}~\cite{Greenblatt:LISP:1980} and FLATS\cite{Goto:Lisp:1982}. Some of these hardware projects have left a lasting impact on modern ways of using personal computers, while others were eclipsed by the rising tide of cheap fast processors.  See the obituary of Eiichi Goto (1931--2005)~\cite{Ida2005}.

\end{itemize}

\subsection{Other histories}
We have mentioned already some historically-oriented papers. Then there is the telegraphic list at 
Brian Evans's History of Computer Algebra\footnote{\url{http://felix.unife.it/Root/d-Mathematics/d-The-mathematician/d-History-of-mathematics/t-History-of-computer-algebra}}.

In~\cite{VELEZ2022} we find a history of computer algebra starting in the 1980s, traced back through the Spanish engineer and inventor Torres--Quevedo (1852--1936) and before.  That paper takes as its actual starting point the results of SYMSAC~'86 and the meeting of the one of the present authors there with Wen-ts\"un Wu, and works its way backward. 

Many other papers attempt at least a brief summary of the history of computer algebra.  One notable such paper is~\cite{decker2001some}, which begins with Babbage and notes several milestones.  

\section{Significant omissions}
We have not discussed the history of the use of computer algebra in education, even though it has been very important for many members of our community since the beginning. We might have included the creation of the computer language \sysname{Logo} in 1967.  Instead we point to the history at~\cite{solomon2020history}, and leave that story for another day.

Automatic reasoning, then called theorem proving, was mentioned as an interest of the members of SICSAM in the very first issue of the Bulletin. Yet we have not mentioned much about the developments in 1965--1975 apart from the quantifier elimination work. There is much more to say.

We mentioned the two-dimensional math output of MathLab, but the whole subject of human-computer interaction has had a ``test laboratory'' in computer algebra systems.  Input of complex objects, control of highly structured objects where transformations from one class to another are routinely desired, and sophisticated visualization of the output have all been desiderata.  Simultaneously there has been a strong need for standardized dissemination of the results.

Neither have we talked about visualization or illustration of mathematical concepts.  One very amusing visualization in the 1975 Bulletin is by Donald~E.~Knuth~\cite{knuth1975seminumerical}, where in describing the then-current state of his ``Volume 2, Seminumerical algorithms'' he includes a hand-drawn chart of editing changes.  

\section{Concluding remarks}
We have given a brief summary of work in computer algebra and symbolic computation, concentrating on the period 1965--1975, ending fifty years ago.  This time period was chosen in part because ISSAC 2025 in Guanajuato, Mexico was the fiftieth since SYMSAC '66, the founding conference organized by Jean Sammet. We have spent a bit more space in this paper in detailing her contributions, which we believe were formative for the community.

During 1965--1975, several important features of our current set of tools were established: efficient methods for dealing with polynomials and series, effective methods for differentiation and integration, methods for solving linear and nonlinear algebraic equations, and manipulations of special functions.
One major idea driving our
subject early on was Artificial
Intelligence. Another could loosely be termed ``solving equations''. A major motivating application was computing formulas in a variety of areas of physics. In each case the
work was notable for being sharply constrained by the amount of memory
that was available.

A half-century and more later, we can see the discipline of ``symbolic computation,'' 
which had split
off into its own island state for a while, beginning to move back to merge with other
sorts of computer work, bringing with it its style of mathematical
discipline. And as it impinges on AI it joins in where hundreds of
gigabytes strain current resources. Application areas that will have
benefited from analytic derivation of the exact computation rules for
computational chemistry and fluid dynamics and others are in a similar situation. 

Jean Sammet observed early on that ``expressions grow to exceed the space
available''~\cite{Sammet:PLHF:1969}. This has held true to a really astonishing extent!
Even today, Gr\"obner Bases and Quantifier Elimination can have
explosive memory demands even on problems of modest size.

It is becoming evident that knowing the history of computer algebra and symbolic computation is important for the general public, not just for us.
In the week previous to ISSAC 2025, some of the authors of this paper ran a session on historical topics at the 30th Applications of Computer Algebra (ACA) conference in Crete.
The plan there was to have a number of first-hand reports from individuals who
were present in earlier days, and then to form a group to build up a fuller
record of this history in book form. That session was a success, and the project is ongoing.  Those who read this paper and who have
relevant information or insight are encouraged to get in touch so that they
can provide input to that enterprise.

\section*{Acknowledgements}
This project was begun in a conversation at a meeting of the Computational Epistemology Think Tank run by The Rotman Institute of Philosophy.
We have relied significantly on the record kept by the ACM Digital Library (it is somewhat ironic, given that this is a paper on the early history of artificial intelligence, that we had to certify so many times when accessing the Library that we were human).
We are grateful to Bruno Buchberger for encouragement and comments on an earlier draft. We also thank Michael Monagan, David Stoutemyer, and the ISSAC referees for comments on the previous version of the paper~\cite{HalfCentury2025}, for all their extremely helpful comments and suggestions. Thanks also to Santiago Recio, for his help dealing with the possible Latin inspiration of the name \sysname{Macsyma}.


\bibliographystyle{plain}
\bibliography{algebra50}
\end{document}